\documentclass[letterpaper]{article} 
\usepackage{aaai23_arxiv}
\usepackage{times}  
\usepackage{helvet}  
\usepackage{courier}  
\usepackage[hyphens]{url}  
\usepackage{graphicx} 
\usepackage{natbib}  
\usepackage{caption} 
\usepackage[utf8]{inputenc} 
\usepackage[T1]{fontenc}    
\usepackage{hyperref}       
\usepackage{booktabs}       
\usepackage{amsfonts}       
\usepackage{nicefrac}       
\usepackage{microtype}      
\usepackage[x11names,dvipsnames]{xcolor}
\usepackage{multirow}

\usepackage{algorithmic}
\usepackage[ruled,linesnumbered]{algorithm2e}

\usepackage{subfig}
\usepackage{amssymb}
\usepackage{mathrsfs}
\usepackage{amsmath}
\usepackage{bm}
\usepackage{colortbl}

\usepackage{bbding}

\usepackage[toc,page,header]{appendix}

\newcommand{\name}{MetaTPTrans}

\title{\name{}: A Meta Learning Approach for Multilingual Code \\Representation Learning}

\author{
    Weiguo Pian\textsuperscript{\rm 1}, Hanyu Peng\textsuperscript{\rm 2}, Xunzhu Tang\textsuperscript{\rm 1}, Tiezhu Sun\textsuperscript{\rm 1}, Haoye Tian\textsuperscript{\rm 1}\thanks{Corresponding author.}, \\
    Andrew Habib\textsuperscript{\rm 1}, Jacques Klein\textsuperscript{\rm 1}, Tegawend{\'{e}} F. Bissyand{\'{e}}\textsuperscript{\rm 1,3}
}
\affiliations{
    \textsuperscript{\rm 1} SnT, University of Luxembourg, Luxembourg \\
    \textsuperscript{\rm 2} Baidu Inc., Beijing, China \\
     \textsuperscript{\rm 3} CITADEL, Université Virtuelle du Burkina Faso
    
    {\rm weiguo.pian@uni.lu \quad penghanyu@baidu.com \quad \{xunzhu.tang, tiezhu.sun, haoye.tian\}@uni.lu \\
    andrew.a.habib@gmail.com \quad \{jacques.klein, tegawende.bissyande\}@uni.lu}
}

\usepackage{bibentry}

\newcommand{\codeSumPrecImprMinPctPt}{0.76\xspace}
\newcommand{\codeSumPrecImprMaxPctPt}{2.38\xspace}

\newcommand{\codeSumRecImprMinPctPt}{1.52\xspace}
\newcommand{\codeSumRecImprMaxPctPt}{2.65\xspace}

\newcommand{\codeSumFOneImprMinPctPt}{1.11\xspace}
\newcommand{\codeSumFOneImprMaxPctPt}{2.40\xspace}

\begin{document}

\maketitle

\begin{abstract}
Representation learning of source code is essential for applying machine learning to software engineering tasks. Learning code representation from a multilingual source code dataset has been shown to be more effective than learning from single-language datasets separately, since more training data from multilingual dataset improves the model's ability to extract language-agnostic information from source code. However, existing multilingual training overlooks the language-specific information which is crucial for modeling source code across different programming languages, while only focusing on learning a unified model with shared parameters among different languages for language-agnostic information modeling. To address this problem, we propose \name, a meta learning approach for multilingual code representation learning. \name{} generates different parameters for the feature extractor according to the specific programming language type of the input code snippet, enabling the model to learn both language-agnostic and language-specific information with dynamic parameters in the feature extractor. We conduct experiments on the code summarization and code completion tasks to verify the effectiveness of our approach. The results demonstrate the superiority of our approach with significant improvements on state-of-the-art baselines.
\end{abstract}


\section{Introduction}

Modeling source code aims at capturing both its syntax and semantics to enable applying machine learning to software engineering tasks such as code summarization~\citep{zugner2021language, peng2021integrating}, code completion~\citep{liu2020self,liu2020multi}, bug finding~\citep{wang2016bugram, pradel2018deepbugs}, patch correctness assessment~\cite{tian2022predicting, tian2022best}, etc. Inspired by the success of deep learning in the field of natural language processing (NLP)~\citep{hochreiter1997long, vaswani2017attention}, modeling source code with deep learning techniques has attracted increasing attention from researchers in recent years~\citep{alon2019code2seq, alon2019code2vec, Hellendoorn2020Global, kim2021code}. 
Although programs are more repetitive and predictable (i.e. "natural"~\citep{hindle2012on}), unlike natural language text, they also contain rich structural information, e.g., ASTs, and data- and control-flow information.
Therefore, a straightforward adoption of NLP models to source code struggles to capture structural information from the source code context (sequence of tokens).

To alleviate the problem above, \citet{alon2019code2seq,alon2019code2vec} proposed to incorporate the structural information by encoding pairwise paths on AST via LSTMs~\citep{hochreiter1997long} to represent the source code.
Inspired by the success of Graph Neural Networks (GNN)~\citep{kipf2017semi,hamilton2017inductive,velickovic2018graph,Gasteiger2020Directional} in modeling structural data, several works leverage GNNs on ASTs to capture the program structure. 
For instance, \citet{allamanis2018learning} used Gated Graph Neural Networks (GGNN) to learn program representation over ASTs and data flow graphs.
Based on GGNN and programs graphs, \citet{fernandes2019structured} proposed Sequence GNN for code summarization and \citet{zhou2019devign} proposed a GNN-based approach to detect software vulnerabilities. However, GNN-based methods struggle to extract global structural information since GNNs focus on local message passing when aggregating information across nodes. Besides, the aforementioned approaches focus on structural information and ignore the contextual information when extracting features from source code. To model both structural and contextual information, \citet{Hellendoorn2020Global} combined the two kinds of information through Transformers~\citep{vaswani2017attention} by biasing the self-attention process with relation information extracted from graph edge types. \citet{zugner2021language} proposed to bias the computation of self-attention with multiple kinds of pairwise distances between AST nodes and integrate them into the XLNet model~\citep{yang2019xlnet}. 
Lastly, \citet{peng2021integrating} introduced TPTrans, which biases the attention processes in Transformers with the encoding of relative and absolute AST paths.

\begin{figure*}
    \centering
    \includegraphics[width=0.68\textwidth,height=0.4\textwidth]{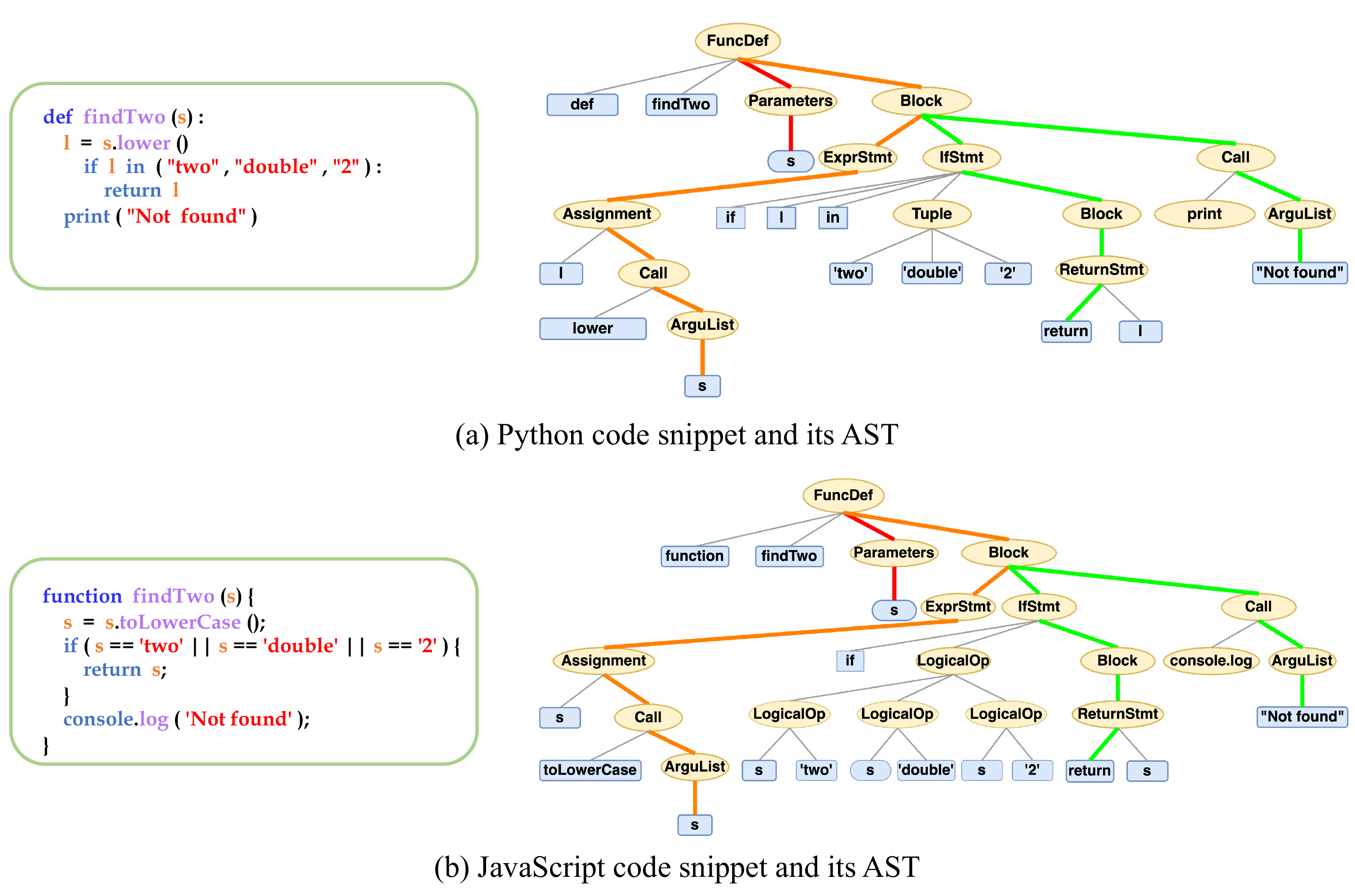}
    \vspace{-8pt}
    \caption{The same code snipped and its associated AST in (a) Python and (b) JavaScript.}
    \label{fig:example}
    \vspace{-8pt}
\end{figure*}

In a different direction, \citet{zugner2021language} introduced their novel insight that multilingual training improves the performance of the model compared to training models on single-language datasets separately, since training the model with multilingual source code data enhances the model's ability to learn language-agnostic information~\citep{zugner2021language}. The language-agnostic information is introduced in~\citep{zugner2021language}, which means that the information can be directly extracted from the source code or AST by a unified model across different languages, without relying on any language-specific features~\citep{zugner2021language}. Such as structural information from the AST of the code, since these ASTs have a unified structural (tree) form across different programming languages (different programming languages have a unified rule of constructing an AST from the source code, and the structural modeling of ASTs is essentially to make the model understand such unified structural rule of them). 
In addition to language-agnostic information, other information is defined as language-specific information, e.g., the language-specific underlying syntactic rules, the language-specific API names, etc. As an example of language-agnostic information, Figure~\ref{fig:example} shows two code snippets of the same function in (a) Python and (b) JavaScript. The two ASTs of the code snippets are shown on the right hand side of the respective figures. 
We can obviously see that the two ASTs have a unified structural (tree) form, besides, we also observe that they have some shared paths, e.g., the highlighted paths in red, orange, and green. Because of such language-agnostic information which can be extracted by a unified model directly across different languages, multilingual training shows superior performance compared to training models on single-language datasets separately.

Although the multilingual training strategy improves the model's performance significantly, it ignores the language-specific information hidden in the characteristics of each specific language.
Considering the same example shown in Figure~\ref{fig:example}, we see that the context of the code snippets (the sequence of source code tokens) in different languages carries distinctive language-specific features, e.g., the different coding rules of the code context in different languages (language-specific underlying syntactic rules), the language-specific API names in different languages:
\texttt{lower()} and \texttt{print()} in Python vs. \texttt{toLowerCase()} and \texttt{console.log()} in JavaScript. For the existing multilingual training, it struggles to learn such language-specific syntactic rules and language-specific token representation/projection with a unified model, \textit{since a unified model is more inclined to learn the common characteristics of the input data}, i.e., the language-agnostic information under the scenario of multilingual code modeling.

To address this problem, in this paper, we propose \name{}, a meta learning based approach for multilingual code representation learning. Meta learning is a novel learning paradigm with the concept of learning to learn. There are several forms of learning to learn in meta learning, such as learning to initialize~\citep{Finn2017MAML}, learning the optimizer~\citep{Andrychowicz2016learning}, learning hyperparameters~\citep{Li2021metasaug}, learning to generate parameters of the feature extractor by another network~\cite{Bertinetto2016learning, Ha2016hyperNetworks,chen2018meta,wang2019tafenet,pan2019urban,pan2022spatio}. \textit{The meta learning form we applied is the last one, i.e., learning to generate the model's parameters, in which the feature extractor can be named as Base Learner while the network for generating parameters is named as Meta Learner}~\citep{wang2019tafenet,pan2019urban,pan2022spatio}.
With this form of meta learning, our model can adjust its parameters dynamically regarding the programming language type of the input code snippet. 
This enables \name{} to not only extract language-agnostic information from multilingual source code data, but also capture the language-specific information, which is overlooked by standard multilingual training. 
Specifically, our approach consists of a language-aware Meta Learner and a feature extractor (Base Learner). The Meta Learner takes the programming language type (e.g., \texttt{Python})
as input and outputs a specific group of parameters for the feature extractor based on the associated language type. The language type is like an indicator to guide the Meta Learner to output specific parameters for a specific programming language.
For the Base Learner, we apply TPTrans~\citep{peng2021integrating}, a state-of-the-art code representation model, as the architecture of it.
To the best of our knowledge, we are the first to leverage meta learning to learn to generate language-aware dynamic parameters for the feature extractor to learn both language-specific and language-agnostic information from multilingual source code datasets.
We evaluate \name{} on two common software engineering tasks: code summarization and code completion.
The experimental results show that our approach outperforms state-of-the-art methods significantly on both tasks.

In summary, this paper contributes the following: 
(1) \name{}, a meta learning based approach for multilingual code representation learning, which learns both language-agnostic and language-specific information from multilingual source code data,
(2) Three different schemes for generating parameters for different kinds of weights of the Base Learner in \name{}, and
(3) Experimental evaluation on two important and widely-used software engineering tasks: code summarization and code completion, and the results show that our approach outperforms state-of-the-art baselines significantly. 
Our code is available at: \url{https://github.com/weiguoPian/MetaTPTrans}.


\section{Technical Preliminaries}
In this section, we introduce the foundations of absolute and relative position embedding in self-attention and the TPTrans model~\citep{peng2021integrating}, upon which we build our model.

\subsection{Self-Attention with Absolute and Relative Position Embedding}
Self-attention (SA) is the basic module in Transformers~\citep{vaswani2017attention}. It maintains three projected matrices $\boldsymbol{Q}\in\mathbb{R}^{d_q\times d_q}$, $\boldsymbol{K}\in\mathbb{R}^{d_k\times d_k}$, and $\boldsymbol{V}\in\mathbb{R}^{d_v\times d_v}$ to compute an output that is the weighted sum of the input by attention score:
\begin{equation}
    \begin{split}
        SA(\boldsymbol{Q},\boldsymbol{K},\boldsymbol{V}) &= softmax(\frac{\boldsymbol{QK}^\text{T}}{\sqrt{d}})\boldsymbol{V} \\
        \textit{s.t.}\quad
        \begin{bmatrix}
		\boldsymbol{Q} \\
		\boldsymbol{K} \\
		\boldsymbol{V}
	    \end{bmatrix}&=\boldsymbol{X}
	    \begin{bmatrix}
		\boldsymbol{W}^Q \\
		\boldsymbol{W}^K \\
		\boldsymbol{W}^V
	    \end{bmatrix}
    \end{split}
\end{equation}
where $\boldsymbol{X}=(\boldsymbol{x}_1,\boldsymbol{x}_2,...,\boldsymbol{x}_n)$ is the input sequence of the self-attention module, $\boldsymbol{x}_i\in\mathbb{R}^{d}$, $d$ is the dimension of the hidden state, and $\boldsymbol{W}^Q\in\mathbb{R}^{d\times d_q}$, $\boldsymbol{W}^K\in\mathbb{R}^{d\times d_k}$, $\boldsymbol{W}^V\in\mathbb{R}^{d\times d_v}$ are the learnable parameters matrices of the self-attention component. Here, we follow the previous works~\citep{vaswani2017attention,zugner2021language,peng2021integrating} and set $d_q=d_k=d_v=d$. More specifically, the above equation can be reformulated as:
\begin{equation}
    \begin{split}
    \boldsymbol{z}_i = \sum\limits_{j=1}^{n}&\frac{\exp(\alpha_{ij})}{\sum_{k=1}^n \exp(\alpha_{ik})}(\boldsymbol{x}_j\boldsymbol{W}^V) \\
    \textit{s.t.}\quad \alpha_{ij} &= \frac{(\boldsymbol{x}_i\boldsymbol{W}^Q)(\boldsymbol{x}_j\boldsymbol{W}^K)^\text{T}}{\sqrt{d}}
    \end{split}
\end{equation}
where $\boldsymbol{z}_i$ is the output of $\boldsymbol{x}_i$ calculated by self-attention operation.
In Vanilla Transformer, \citet{vaswani2017attention} used a non-parameteric absolute position encoding, which is added to the word vectors directly.~\citet{ke2021rethinking} proposed a learnable projection for absolute position for computing the attention score among words:
\begin{equation}
    \begin{split}
        \alpha_{ij}=\frac{(\boldsymbol{x}_i\boldsymbol{W}^Q)(\boldsymbol{x}_j\boldsymbol{W}^K)^\text{T}}{\sqrt{2d}} + \frac{(\boldsymbol{p}_i\boldsymbol{U}^Q)(\boldsymbol{p}_j\boldsymbol{U}^K)^\text{T}}{\sqrt{2d}}
    \end{split}
    \label{eq:absolute_pos}
\end{equation}
where $\boldsymbol{p}_i$ denotes the learnable real-valued vector of position $i$, and $\boldsymbol{U}^Q,\boldsymbol{U}^K\in \mathbb{R}^{d\times d}$ are the projection matrices of the position vectors $\boldsymbol{p}_i$ and $\boldsymbol{p}_j$ respectively. To capture the relative position relationship between words, \citet{shaw2018self} proposed to use the relative position embedding between each two words:
\begin{equation}
\begin{split}
    \boldsymbol{z}_i = \sum\limits_{j=1}^{n}&\frac{\exp(\alpha_{ij})}{\sum_{k=1}^n \exp(\alpha_{ik})}(\boldsymbol{x}_j\boldsymbol{W}^V+\boldsymbol{r}_{ij}^V) \\
    \textit{s.t.}\quad \alpha_{ij} &= \frac{(\boldsymbol{x}_i\boldsymbol{W}^Q)(\boldsymbol{x}_j\boldsymbol{W}^K+\boldsymbol{r}_{ij}^K)^\text{T}}{\sqrt{d}}
    \label{eq:relative_pos}
\end{split}
\end{equation}
where $\boldsymbol{r}_{ij}^K,\boldsymbol{r}_{ij}^V$ denote the learnable relative position embedding between positions $i$ and $j$.

\subsection{TPTrans}
We briefly introduced how absolute and relative position embeddings are integrated into the self-attention module of Transformers. In this subsection, we describe the TPTrans~\citep{peng2021integrating} which is based on the aforementioned position embedding concepts.

TPTrans modifies the relative and absolute position embedding in self-attention with AST paths encodings so that AST paths could be integrated into Transformers. Specifically, they first encode the relative and absolute path via a bi-directional GRU~\citep{cho2014properties}:
\begin{equation}
    \begin{split}
    \boldsymbol{r}_{ij} &= GRU(\boldsymbol{Path}_{\boldsymbol{x}_i\rightarrow\boldsymbol{x}_j}) \\
    \boldsymbol{a}_i &= GRU(\boldsymbol{Path}_{\boldsymbol{root}\rightarrow\boldsymbol{x}_i})
    \end{split}
\end{equation}
where $\boldsymbol{Path}_{\boldsymbol{x}_i\rightarrow\boldsymbol{x}_j}$ denotes the AST path from node $\boldsymbol{x}_i$ to node $\boldsymbol{x}_j$, $\boldsymbol{r}_{ij}$ is the relative path encoding between positions $i$ and $j$, and $\boldsymbol{a}_i$ is the absolute path (from the $\boldsymbol{root}$ node to node $\boldsymbol{x}_i$) encoding of position $i$. 
Then, the two types of path encodings are integrated into Eq.~(\ref{eq:absolute_pos}) -~(\ref{eq:relative_pos}) by replacing the absolute and relative position embeddings:
\begin{equation}
\begin{split}
    \boldsymbol{z}_{i}=\sum\limits_{j=1}^n&\frac{\exp(\alpha_{ij})}{\sum_{k=1}^n\exp(\alpha_{ik})}(\boldsymbol{x}_j\boldsymbol{W}^V+\boldsymbol{r}_{ij}\boldsymbol{W}^{r,V}) \\
    \textit{s.t.}\quad \alpha_{ij} &= \frac{(\boldsymbol{x}_i\boldsymbol{W}^Q) (\boldsymbol{x}_j\boldsymbol{W}^K+\boldsymbol{r}_{ij}\boldsymbol{W}^{r,K})^\text{T}}{\sqrt{d}} \\
    &+ \frac{(\boldsymbol{a}_i\boldsymbol{W}^{a,Q})(\boldsymbol{a}_j\boldsymbol{W}^{a,K})^\text{T}}{\sqrt{d}}
    \label{eq:TPTrans}
\end{split}
\end{equation}
where $\boldsymbol{W}^{r,K},\boldsymbol{W}^{r,V}\in\mathbb{R}^{d\times d}$ are the key and value projection matrices of the relative path encoding and $\boldsymbol{W}^{a,Q},\boldsymbol{W}^{a,K}$ denote the query and key projection matrices of the absolute path encoding.


\section{Approach}
\begin{figure*}
    \centering
    \includegraphics[width=0.72\textwidth]{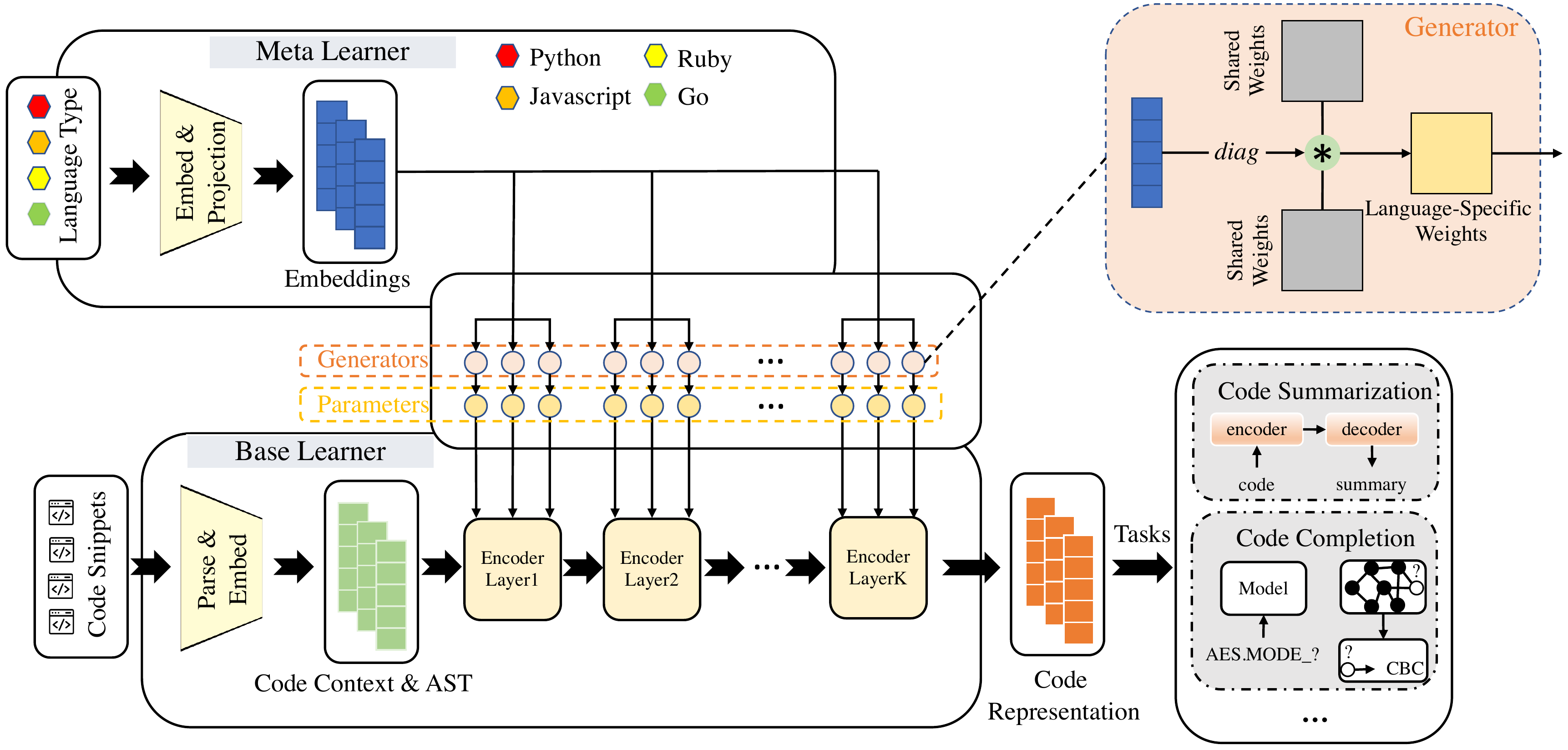}
    \caption{Architecture of \name{}.
    }
    \label{fig:overview}
    \vspace{-6pt}
\end{figure*}

We introduce \name{} which consists of two components: a Meta Learner and a Base Learner. Specifically, the Meta Learner takes the language type as input and generates language-specific parameters for the Base Learner. We apply the TPTrans as the architecture of the Base Learner, which takes the source code as input, and uses the language-specific parameters generated from the Meta Learner to calculate the representation of the input source code snippet. The overview of our approach is shown in Figure~\ref{fig:overview}.

\subsection{Meta Learner}
The Meta Learner generates parameters for the Base Learner according to the language type. Specifically, for a given source code snippet $\boldsymbol{X}_i$ and its corresponding language type $t_i$, the \textit{Language Embedding Layer} $\mathcal{T}$ embeds the language type $t_i$ into an embedding $\boldsymbol{T}_i\in\mathbb{R}^{d_T}$. Then, a projection layer scales the dimension of $\boldsymbol{T}_i$. The overall process can be presented as:
\begin{equation}
    \begin{split}
    &\boldsymbol{T}_i=\mathcal{T}(t_i) \\
    \boldsymbol{P}_i=&Projection(\boldsymbol{T}_i)
    \end{split}
\end{equation}
where $Projection(\cdot)$ and $\boldsymbol{P}_i\in\mathbb{R}^{d_P}$  denote the projection layer and the projected language type embedding, respectively.
After producing the projected language type embedding $\boldsymbol{P}_i$, we now present the parameter generation scheme for the Base Learner. For a weight matrix $\boldsymbol{W}^\lambda\in\mathbb{R}^{d\times d}$ in the parameters of Base Learner, we conduct it via a generator $\mathcal{G}_\lambda$, which can be denoted as:
\begin{equation}
    \boldsymbol{W}^\lambda=\mathcal{G}_\lambda(\boldsymbol{P}_i)
\end{equation}
As noted in~\citep{Bertinetto2016learning,wang2019tafenet}, using a linear projection layer to scale a $d_P$-dimension vector to a matrix of dimension $d\times d$ exhibits a large memory footprint, especially that there are many such weight matrices in the parameters of the Base Learner. To reduce the computational and memory cost, we adopt the factorized scheme for the weight generation procedure by factorizing the representation of the weights, which is analogous to the \textit{Singular Value Decomposition}~\citep{Bertinetto2016learning}.
In this way, the original vector can be projected into the target matrix dimension space with fewer parameters in the generator. 
More specifically, we first apply the diagonal operation to transform the vector $\boldsymbol{P}_i$ to a diagonal matrix, then two projection matrices $\boldsymbol{M}_\lambda\in\mathbb{R}^{d_P\times d}$ and $\boldsymbol{M'}_\lambda\in\mathbb{R}^{d\times d_P}$ are defined to project the diagonal projected language type embedding matrix into the space of the target weight matrix. This operation can be formulated as:
\begin{equation}
    \boldsymbol{W}^\lambda=\mathcal{G}_\lambda(\boldsymbol{P}_i)=\boldsymbol{M'}_{\lambda}diag(\boldsymbol{P}_i)\boldsymbol{M}_\lambda
\end{equation}
where $diag(\cdot)$ is the non-parametric diagonal operation, and $\boldsymbol{M}_\lambda$ and $\boldsymbol{M'}_\lambda$ are the learnable parameters in generator $\mathcal{G}_\lambda$.



\subsection{Base Learner}

The Base Learner is the module that actually learns the representation of code snippets, the parameters of which are generated from the Meta Learner.
In our approach, we apply the TPTrans model as the architecture of the Base Learner. Recall from Eq.~(\ref{eq:TPTrans}), the learnable parameters of TPTrans consist of the projection matrices of tokens ($\boldsymbol{W}^Q,\boldsymbol{W}^K,\boldsymbol{W}^V$) and the projection matrices of path encodings ($\boldsymbol{W}^{r,K}, \boldsymbol{W}^{r,V}, \boldsymbol{W}^{a,Q}, \boldsymbol{W}^{a,K}$). 

The Meta Learner generates the parameters of the Base Learner. First, we consider that the most obvious language-specific information is the language-specific underlying syntax rule in the context of the code snippet. Such contextual information is extracted from the input token sequences. Therefore, we first generate the projection matrices for token sequences. This procedure can be formulated as:
\begin{equation}
    \begin{split}
        \boldsymbol{W}^\lambda_{t_i}=\mathcal{G}_\lambda(&\boldsymbol{P}_i)=\boldsymbol{M'}_\lambda diag(\boldsymbol{P}_i)\boldsymbol{M}_\lambda \\
        \textit{s.t.}\;\; &\lambda\in\{Q,K,V\}
        \label{eq:generator_context}
    \end{split}
\end{equation}
where $t_i$ denotes the corresponding language type of code snippet $\boldsymbol{X}_i$, and $\boldsymbol{W}^Q_{t_i},\boldsymbol{W}^K_{t_i},\boldsymbol{W}^V_{t_i}$ are the learnable weights of $\boldsymbol{W}^Q,\boldsymbol{W}^K,\boldsymbol{W}^V$ associated with language type $t_i$, and $\boldsymbol{P}_i=Projection(\mathcal{T}(t_i))$ denotes the projected language type embedding of $t_i$. After that, the generated weights matrices are assigned to the related parameters by replacing the unified related weights matrices in Eq.~(\ref{eq:TPTrans}):
\begin{equation}
\begin{split}
    \boldsymbol{z}_{i}=\sum\limits_{j=1}^n&\frac{\exp(\alpha_{ij})}{\sum_{k=1}^n\exp(\alpha_{ik})}(\boldsymbol{x}_j \mathcal{G}_V(\boldsymbol{P}_i) +\boldsymbol{r}_{ij}\boldsymbol{W}^{r,V}) \\
    \textit{s.t.}\quad \alpha_{ij} &= \frac{(\boldsymbol{x}_i\mathcal{G}_Q(\boldsymbol{P}_i)) (\boldsymbol{x}_j\mathcal{G}_K(\boldsymbol{P}_i)+\boldsymbol{r}_{ij}\boldsymbol{W}^{r,K})^\text{T}}{\sqrt{d}} \\
    &+ \frac{(\boldsymbol{a}_i\boldsymbol{W}^{a,Q})(\boldsymbol{a}_j\boldsymbol{W}^{a,K})^\text{T}}{\sqrt{d}}
    \label{eq:MetaTPTrans_alpha}
\end{split}
\end{equation}
Further, the weights matrices for path encoding projection can also be generated by the Meta Learner for different language types. This process aims to integrate the language-agnostic structural information (path encodings) into the contextual information dynamically according to the associated language type, which we believe is meaningful to investigate since the combination of these two kinds of information may also be influenced by the underlying language type.
Similar to Eq.~(\ref{eq:generator_context}), this procedure can be expressed as:
\begin{equation}
    \begin{split}
        \boldsymbol{W}^\lambda_{t_i}=\mathcal{G}&_\lambda(\boldsymbol{P}_i)=\boldsymbol{M'}_\lambda diag(\boldsymbol{P}_i)\boldsymbol{M}_\lambda \\
        \textit{s.t.}\;\; \lambda\in\{\{r&,K\}, \{r,V\}, \{a,Q\}, \{a,K\} \}
        \label{eq:generator_structure}
    \end{split}
\end{equation}
where $\boldsymbol{W}^{r,K}_{t_i},\boldsymbol{W}^{r,V}_{t_i},\boldsymbol{W}^{a,Q}_{t_i},\boldsymbol{W}^{a,K}_{t_i}$ are the generated weights of $\boldsymbol{W}^{r,K},\boldsymbol{W}^{r,V},\boldsymbol{W}^{a,Q},\boldsymbol{W}^{a,K}$ associated with language type $t_i$. 
After that, the generated weights matrices in Eq.~(\ref{eq:generator_structure}) can be integrated into Eq.~(\ref{eq:TPTrans}) by replacing the related weights matrices:
\begin{equation}
\begin{split}
    \boldsymbol{z}_{i}=\sum\limits_{j=1}^n&\frac{\exp(\alpha_{ij})}{\sum_{k=1}^n\exp(\alpha_{ik})}(\boldsymbol{x}_j\boldsymbol{W}^V +\boldsymbol{r}_{ij}\mathcal{G}_{r,V}(\boldsymbol{P}_i)) \\
    \textit{s.t.}\quad \alpha_{ij}& = \frac{(\boldsymbol{x}_i\boldsymbol{W}^Q) (\boldsymbol{x}_j\boldsymbol{W}^K+\boldsymbol{r}_{ij}\mathcal{G}_{r,K}(\boldsymbol{P}_i))^\text{T}}{\sqrt{d}} \\
     &+ \frac{(\boldsymbol{a}_i \mathcal{G}_{a,Q}(\boldsymbol{P}_i))(\boldsymbol{a}_j\mathcal{G}_{a,K}(\boldsymbol{P}_i))^\text{T}}{\sqrt{d}}
    \label{eq:MetaTPTrans_beta}
\end{split}
\end{equation}
Finally, we combine the above two kinds of weights generation schemes in which both context token projection and path encoding projection are generated by the Meta Learner:
\begin{equation}
\begin{split}
    \boldsymbol{z}_{i}=\sum\limits_{j=1}^n&\frac{\exp(\alpha_{ij})}{\sum_{k=1}^n\exp(\alpha_{ik})}(\boldsymbol{x}_j \mathcal{G}_V(\boldsymbol{P}_i) +\boldsymbol{r}_{ij}\mathcal{G}_{r,V}(\boldsymbol{P}_i)) \\
    \textit{s.t.}\quad \alpha_{ij} &= \frac{(\boldsymbol{x}_i\mathcal{G}_Q(\boldsymbol{P}_i)) (\boldsymbol{x}_j\mathcal{G}_K(\boldsymbol{P}_i)+\boldsymbol{r}_{ij}\mathcal{G}_{r,K}(\boldsymbol{P}_i))^\text{T}}{\sqrt{d}} \\
    &+ \frac{(\boldsymbol{a}_i \mathcal{G}_{a,Q}(\boldsymbol{P}_i))(\boldsymbol{a}_j\mathcal{G}_{a,K}(\boldsymbol{P}_i))^\text{T}}{\sqrt{d}}
    \label{eq:MetaTPTrans_gamma}
\end{split}
\end{equation}
In the above, we generate weights matrices from three perspectives: 
(i) For context token projection (Eq.~(\ref{eq:MetaTPTrans_alpha})), 
(ii) For path encoding projection (Eq.~(\ref{eq:MetaTPTrans_beta})), and
(iii) For both context token projection and path encoding projection (Eq.~(\ref{eq:MetaTPTrans_gamma})). 
We name those three schemes of weights matrices generation: \textit{\name-$\alpha$}, \textit{\name-$\beta$}, and \textit{\name-$\gamma$}, respectively.

\section{Experimental Setup and Results}
In this section, we present our experimental setup and results.
We evaluate \name{} on two common and challenging tasks: code summarization and code completion. 
For the number of parameters and training time cost compared with the baseline (TPTrans), we present them in the Appendix.




\paragraph{Code Summarization}
Code summarization aims at describing the functionality of a piece of code in natural language and it demonstrates the capability of the models in capturing the semantics of source code~\citep{alon2019code2vec,zugner2021language,peng2021integrating}. 
Similar to previous work~\citep{zugner2021language,peng2021integrating}, we consider a complete method body as the source code input and the method name as the target prediction (i.e. the NL summary) while predicting the method name as a sequence of subtokens. Following~\citet{zugner2021language,peng2021integrating}, we evaluate the performance of our approach and baselines on the code summarization task using the metrics of precision, recall, and F1 scores over the target sequence.

\begin{figure}
    \centering
    \includegraphics[width=0.35\textwidth]{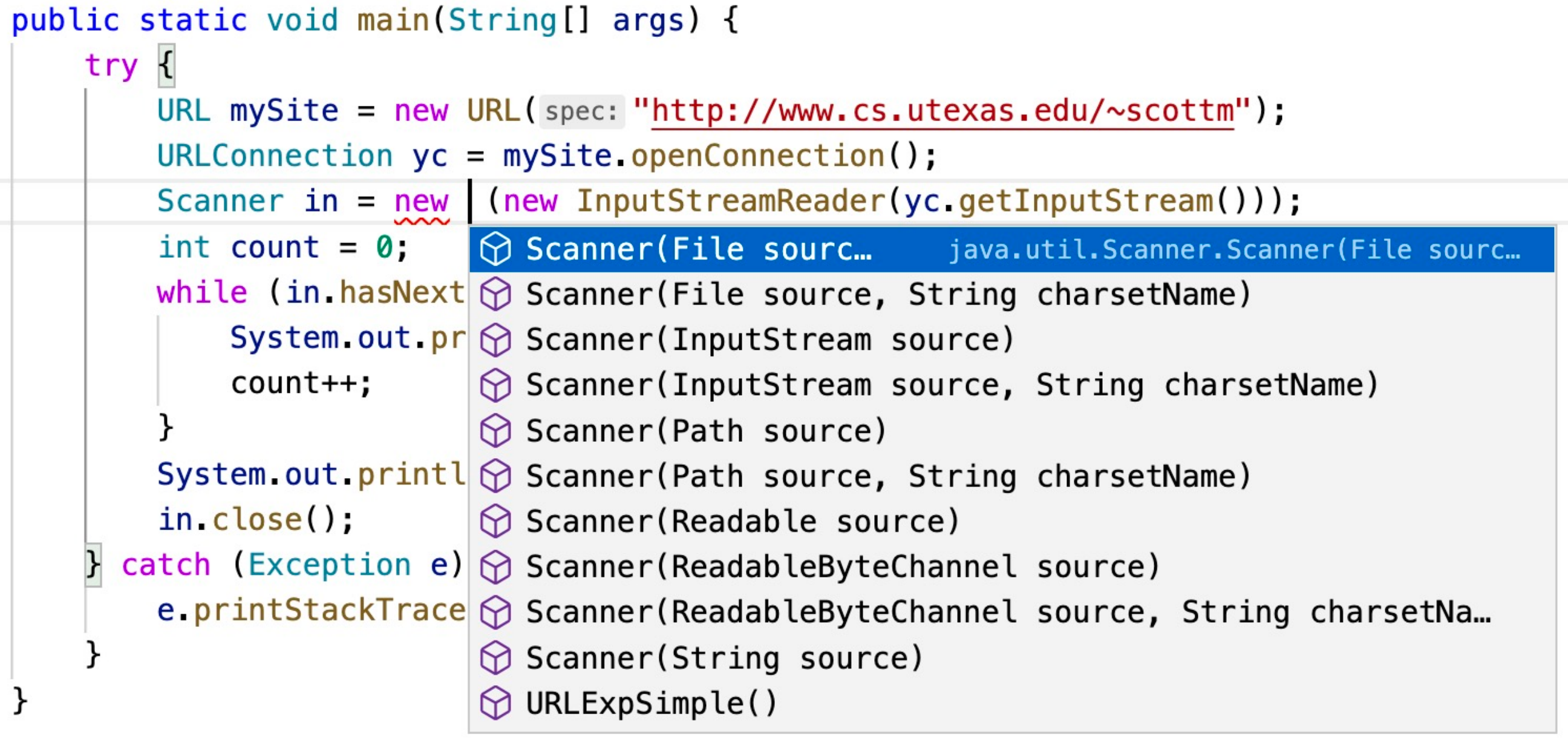}
    \caption{An example of code completion}
    \label{fig:example_completion}
    \vspace{-12pt}
\end{figure}

\paragraph{Code Completion}
Code completion is another challenging downstream task in source code modeling. As the settings in~\citep{liu2020self,liu2020multi}, the model aims to predict a missing token by taking the incomplete code snippet as input.
An example of code completion in realistic scenario is shown in Figure~\ref{fig:example_completion}, in which the IDE predicts a list of possible tokens for a missing token in an incomplete code snippet.
To generate data for code completion task, we randomly replace a token with a special token \texttt{<MASK>} in a given code snippet. And then, the new code snippet with the special token \texttt{<MASK>} is used to predict the replaced missing token.
We report Top-1 and Top-5 prediction accuracy as the evaluation metrics for the code completion task.

\paragraph{Dataset}
We conduct our experiments on the CodeSearchNet~\citep{husain2019codesearchnet} dataset. Following~\citet{zugner2021language,peng2021integrating}, we consider four programming languages in the dataset: Python, Ruby, JavaScript, and Go. 
Please see the Appendix for the details and the pre-processing of the dataset.

\paragraph{Baselines}
For code summarization, we compare \name{} against code2seq~\citep{alon2019code2seq}, GREAT~\citep{Hellendoorn2020Global}, CodeTransformer~\citep{zugner2021language} and TPTrans~\citep{peng2021integrating}. 
For code completion, we compare \name{} against Transformer~\citep{vaswani2017attention}, CodeTransformer~\citep{zhou2019devign} and TPTrans~\citep{peng2021integrating}. Please see the Appendix for more details. 



\begin{table*}[htbp]
  \centering
  \caption{Precision, recall, and F1 for the code summarization task. The bold part denotes the overall best results, and the underlined part denotes the best results of baselines.
  }
    \resizebox{\textwidth}{!}{
    \begin{tabular}{lccc|ccc|ccc|ccc}
    \toprule
    \multirow{2}{*}{Model} & \multicolumn{3}{c|}{Python} & \multicolumn{3}{c|}{Ruby} & \multicolumn{3}{c|}{JavaScript} & \multicolumn{3}{c}{Go} \\
    \cmidrule{2-13} 
    & Prec. & Rec.  & F1    & Prec. & Rec.  & F1    & Prec. & Rec.  & F1    & Prec. & Rec.  & F1 \\
    \midrule
    code2seq (Single-language) & 35.79 & 24.85 & 29.34 & 23.23 & 10.31 & 14.28 & 30.18 & 19.88 & 23.97 & 52.30  & 43.43 & 47.45 \\
    GREAT (Single-language) & 35.07 & 31.59 & 33.24 & 24.64 & 22.23 & 23.38 & 31.20  & 26.84 & 28.86 & 50.01 & 46.51 & 48.20 \\
    CodeTransformer (Single.) & 36.40  & 33.66 & 34.97 & 31.42 & 24.46 & 27.50  & 35.06 & 29.61 & 32.11 & 55.10  & 48.05 & 51.34 \\
    TPTrans (Single-language) & 38.39 & 34.70  & 36.45 & 33.07 & 28.34 & 30.52 & 33.68 & 28.95 & 31.14 & 55.67 & 51.31 & 53.39 \\
    \midrule
    code2seq (Multilingual) & 34.49 & 25.49 & 29.32 & 23.97 & 17.06 & 19.93 & 31.62 & 22.16 & 26.06 & 52.70  & 44.36 & 48.17 \\
    GREAT (Multilingual) & 36.75 & 31.54 & 33.94 & 30.05 & 24.33 & 26.89 & 33.58 & 27.78 & 30.41 & 52.65 & 48.30  & 50.38 \\
    CodeTransformer (Multi.) & 38.89 & 33.82 & 36.18 & 33.93 & 28.94 & 31.24 & \underline{36.95} & 29.98 & \underline{33.10}  & 56.00    & 50.44 & 53.07 \\
    TPTrans (Multilingual) & \underline{39.71} & \underline{34.66} & \underline{37.01} & \underline{39.51} & \underline{32.31} & \underline{35.55} & 34.92  & \underline{30.01} & 32.33 & \underline{56.48} & \underline{52.02} & \underline{54.16} 

    \\

    \midrule
    
     \name-$\alpha$  & 40.22  & \textbf{36.22} & \textbf{38.12} & \textbf{40.62} & \textbf{34.01} & \textbf{37.02} & 37.87 & 31.92 & 34.64 & 58.12 & 53.82 & 55.89 \\
    
     \name-$\beta$ & 39.97 & 36.12 & 37.94 & 40.44 & 33.69 & 36.76 & \textbf{38.87} & \textbf{32.66} & \textbf{35.50} & \textbf{58.86} & \textbf{54.24} & \textbf{56.45} \\
    
     \name-$\gamma$ & \textbf{40.47} & 35.19 & 37.65 & 40.58 & 32.04 & 35.81 & 37.90 & 30.11 & 33.56 & 58.20 & 53.38 & 55.68 \\
    
    \bottomrule
    \end{tabular}%
    }
  \label{tab:res_summarization}%
  \vspace{-2pt}
\end{table*}%

\paragraph{Implementation Details}
For both tasks, following~\citet{peng2021integrating}, we set the embedding sizes of the word, path node, and hidden size of the Transformer to 512, 64, and 1024, respectively. A linear layer projects the word embedding into the size of the hidden layer of the Transformer. We use one bidirectional-GRU~\citep{cho2014properties} layer of size 64 to encode the paths, and concatenate the final states of both directions as output. We use the Adam~\citep{kingma2015adam} optimizer with a learning rate of $1e^{-4}$. 
We train our models for 10 and 40 epochs for the code summarization and code completion tasks, respectively on 4 Tesla V100 GPUs with batch size of 128 and dropout of 0.2.
For the Base Learner in the code summarization task, we use the same hyperparameters setting of TPTrans~\citep{peng2021integrating} for a fair comparison.
Specifically, we set the number of encoder and decoder layers to 3 and 8, the number of attention heads to 3, and the dimension of the feed-forward layer to 4096. 
In the Meta Learner, the dimension of the language type embedding ($d_T$) and its projection  ($d_P$) are set to 1024 and 2048, respectively. Following~\citet{zugner2021language,peng2021integrating}, we add the pointer network~\citep{vinyals2015pointer} to the decoder.
For the code completion task, we set the number of encoder layers, number of heads, and the dimension of feed-forward layers to 5, 8 and 2048 respectively for all the baselines and our approaches.
As we mentioned above, we follow the code completion settings in~\cite{liu2020self,liu2020multi} that predict a missing token amid a incomplete code snippet, which is not a sequence-to-sequence task. Thus, we apply a fully connected layer after the encoder rather than using a decoder to generate the predicted token. In the Meta Learner, we set both the dimension of the language type embedding ($d_T$) and its projection ($d_P$) to 512.

\begin{table*}[t]
  \centering
  \caption{Top-1 and Top-1 accuracy for the code completion task. Underlined, bold values denote the best results in the baselines and the overall best results respectively.
}
    \begin{tabular}{lcc|cc|cc|cc}
    \toprule
    \multirow{2}{*}{Model} & \multicolumn{2}{c|}{Python} & \multicolumn{2}{c|}{Ruby} & \multicolumn{2}{c|}{JavaScript} & \multicolumn{2}{c}{Go} \\
    \cmidrule{2-9} 
    & Top-1  & Top-5  & Top-1  & Top-5  & Top-1  & Top-5  & Top-1  & Top-5  \\
    \midrule
    Transformer (Single-language) & 47.57 & 69.86 & 44.39 & 62.24 & 37.57 & 53.15 & 40.21 & 59.65 \\
    CodeTransformer (Single.) & 62.45 & 76.73  & 51.63 & 69.96 & 47.56 & 68.88 & 47.71 & 61.35 \\
    TPTrans (Single-language) & 63.71 & 77.99 & 64.42 & 72.50 & 64.67 & 73.42 & 57.15 & 67.81 \\
    
    \midrule
    Transformer (Multilingual) & 47.02 & 78.82 & 47.16 & 77.32 & 38.77 & 70.84 & 42.01 & 72.95 \\
    CodeTransformer (Multi.) & 68.19 & 82.98 & 67.67 & \underline{83.47} & 59.32 & 80.07 & 57.12 & 77.67 
    \\
    TPTrans (Multilingual) & \underline{69.81} & \underline{84.10} & \underline{72.14} & 82.27 & \underline{67.45} & \underline{81.17} & \underline{60.45} & \underline{79.03} \\
    
    \midrule
    
     \name-$\alpha$ & \textbf{77.13} & \textbf{94.28} & \textbf{78.05} & \textbf{95.42} & \textbf{73.52} & \textbf{92.88} & \textbf{67.47} & \textbf{91.15} \\
    
     \name-$\beta$ & 71.75 & 86.26 & 73.82 & 86.85 & 72.71 & 86.90 & 66.74 & 85.21 \\
    
     \name-$\gamma$ & 67.12 & 90.72 & 71.89 & 93.61 & 69.55 & 90.97 & 61.60 & 88.99 \\
    
    \bottomrule
    \end{tabular}%
  \label{tab:res_completion}%
  \vspace{-2pt}
\end{table*}%


\subsection{Code Summarization}
Table~\ref{tab:res_summarization} shows the results of the code summarization task. The top part of the table shows the results of the baselines trained on single-language datasets. The middle and bottom parts of the table show the results of the baselines trained on multilingual dataset and our \name{} respectively. \name-$\alpha$ (Eq.~\ref{eq:MetaTPTrans_alpha}), \name-$\beta$ (Eq.~\ref{eq:MetaTPTrans_beta}) and \name-$\gamma$ (Eq.~\ref{eq:MetaTPTrans_gamma}) outperform all the baseline methods significantly.
Specifically, compared with the state-of-the-art results on the Python, Ruby, JavaScript, and Go datasets, our approach improves the F1 score by 1.11, 1.47, 2.40 and 2.29 respectively.
Overall, \name{} improves precision by \codeSumPrecImprMinPctPt--\codeSumPrecImprMaxPctPt, recall by \codeSumRecImprMinPctPt--\codeSumRecImprMaxPctPt, and F1 by \codeSumFOneImprMinPctPt--\codeSumFOneImprMaxPctPt across the four programming languages. 
For the experiment results without pointer networks, please see Appendix for details.

\subsection{Code Completion}
Table~\ref{tab:res_completion} shows the results for the code completion task where the top and middle parts denote the testing results of the baselines trained on single-language datasets and multilingual dataset respectively. And bottom part of the table demonstrates the results of our \name{}. \name-$\alpha$ improves the Top-1 (Top-5) prediction accuracy over the best baseline, by 7.32 (10.18), 5.91 (11.95), 6.07 (11.71) and 7.02 (12.12) for Python, Ruby, JavaScript, and Go, respectively. Among the three variants of \name{}, the \name-$\alpha$ consistently achieves the best results over the four programming languages. As described before, the language-specific weights matrices generated by the Meta Learner in \name-$\alpha$ are only assigned to the parameters of the code context token projection in the Base Learner (Eq.~\ref{eq:MetaTPTrans_alpha}) while \name-$\beta$ and \name-$\gamma$ assign the generated weights to AST path encodings. This means that for code completion, compared with generating language-specific weights for the projection of structural information, generating weights for context token projection is more conducive for the extraction of language-specific information, which leads to a significant performance improvement.

\begin{figure}
\centering
\includegraphics[width=0.44\textwidth]{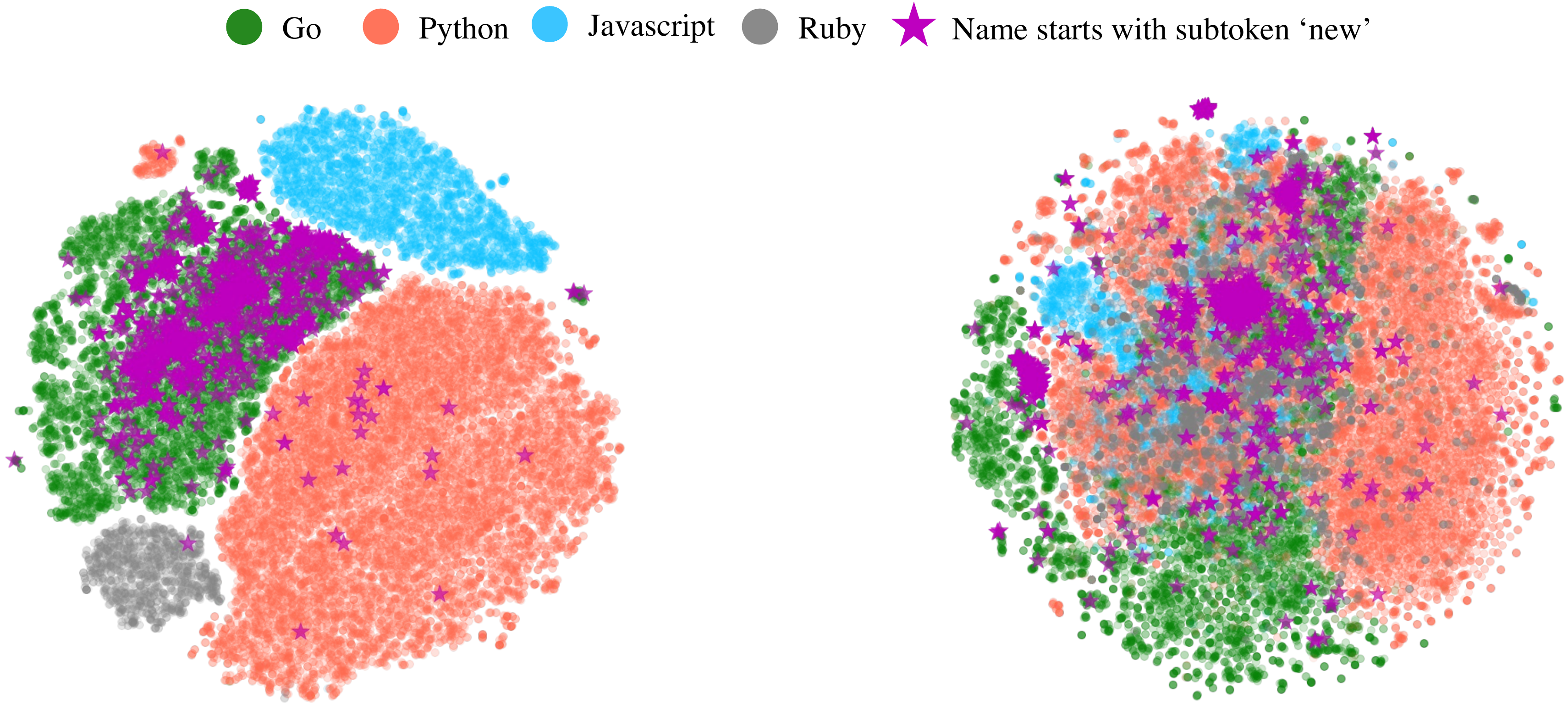}
\caption{t-SNE visualization of the representation learned by \name-$\alpha$ (left) and TPTrans (right).
}
\label{fig:vis}
\vspace{-10pt}
\end{figure}

\subsection{Visualization of Learned Representation}
In figure~\ref{fig:vis}, we show the t-SNE~\citep{van2008visualizing} visualization of the learned representations of all the code snippets from the validation set of the code summarization task. 
The left and right parts of Figure~\ref{fig:vis} show the code representation generated by the encoder of \name-$\alpha$ and the best baseline TPTrans respectively. 
We see that \name-$\alpha$ learns a distributed code representation that also respects the type of programming language of the code snippet as data points from the same language group together.
This demonstrates that our model learns languages-specific features much better than the baseline model, where code snippets from the same language do not necessarily group together.
Moreover, as an example, we mark all the code snippets whose names start with the subtoken \texttt{new} by the symbol \textcolor{Magenta3}{\small\FiveStar}.
We see that \name-$\alpha$ achieves a much better grouping of those code snippets compared to the baseline model emphasizing our model ability to learn a better semantic representation of source code, while also respecting the language-specific features.



\section{Related Work}
\paragraph{Learning Representation of Source Code}
Source code representation learning has seen many developments.
Early works mainly focus on learning language models from raw token sequences~\citep{wang2016bugram,dam2016deep,allamanis2016aconvolutional,iyer2016summarizing}.
More recent works explore the effectiveness of leveraging structural information to model source code. 
\citet{mou2016convolutional} apply the convolutional operation on ASTs to extract structure features to represent source code. \citet{alon2019code2vec,alon2019code2seq} extract paths from ASTs and use RNNs to encode them to represent the source code. 
\citet{allamanis2018learning,fernandes2019structured,zhou2019devign} use Graph Neural Networks to capture the structural information from carefully designed code graphs. 
\citet{Hellendoorn2020Global,zugner2021language,peng2021integrating} use Transformer-based models to represent source code by capturing both context and structural information, in which the structural information is integrated into the self-attention module by replacing the position embedding with encoding from AST. Specifically, \citet{Hellendoorn2020Global} bias the self-attention process with different types of correlations between nodes in code graph. 
\citet{zugner2021language} use several pair-wise distances on ASTs to represent the pair-wise relationships between tokens in code context sequence and find that multilingual training improves the performance of language models compared to single-language models. \citet{peng2021integrating} encode AST paths and integrate them into self-attention to learn both context and structural information.
Compared to these works, ours is the first to use meta learning to learn multilingual source code models that are capable of learning language-specific in addition to language-agnostic information and improves on several of the aforementioned models yielding state-of-the-art results.

\paragraph{Meta Learning for Parameters Generation}
Meta learning is a novel learning paradigm with the concept of learning to learn. There are several types of meta learning such as 
learning to initialize~\citep{Finn2017MAML}, 
learning the optimizer~\citep{Andrychowicz2016learning}, and
learning hyperparameters~\citep{Li2021metasaug}. The most related meta learning method with ours is learning to generate model's parameters. \citet{Bertinetto2016learning} propose a method called learnet to learn to generate the parameters of the pupil network. \citet{Ha2016hyperNetworks} propose Hypernetworks to generate parameters for large models through layer-wise weight sharing scheme. \citet{chen2018meta} propose a meta learning-based multi-task learning framework in which a meta network generates task-specific weights for different tasks to extract task-specific semantic features.
\citet{wang2019tafenet} use a task-aware meta learner to generate parameters for classification models for different tasks in few-shot learning. \citet{pan2019urban,pan2022spatio} apply meta learning to generate parameters for models in spatial-temporal data mining to capture spatial and temporal dynamics in urban traffic. 
Our approach is a form of meta learning to generate the model's parameters according to the programming language type of the input code snippet.

\section{Conclusion}
We propose \name{}, a meta learning approach for multilingual code representation learning. 
Instead of keeping the feature extractor with a fixed set of parameters across different languages, we adopt meta learning to generate different sets of parameters for the feature extractor according to the language type of the input code snippet. 
This enables \name{} to not only extract language-agnostic information, but to also capture language-specific features of source code. 
Experimental results show that our approach outperforms the state-of-the-art baselines significantly.
Our work provides a novel direction for multilingual code representation learning.

\section{Acknowledgments}
This work is supported by the NATURAL project, which has received funding from the European Research Council (ERC) under the European Union’s Horizon 2020 research and innovation programme (grant No. 949014).


\bibliography{aaai23}




\newpage
\appendix

\section{Appendix}



\subsection{Summary of the CodeSearchNet Dataset}
\label{summary_dataset}

We show the summary of the CodeSearchNet~\citep{husain2019codesearchnet} dataset used in our experiments in Table~\ref{tab:dataset}, which contains four programming languages: Python, Ruby, JavaScript, and Go. The dataset has been deduplicated by the creators to avoid data leakage from training set. 


  \begin{table}[htbp]
  \centering
  \caption{Dataset statistics}
    \begin{tabular}{lccc}
    \toprule
    \multirow{2}{*}{Language} & \multicolumn{3}{c}{Samples per partition} \\
    \cmidrule{2-4}          
    & Train & Valid & Test \\
    \midrule
    Python & 412,178 & 23,107 & 22,176 \\
    Ruby & 48,791 & 2,209 & 2,279 \\
    JavaScript & 123,889 & 8,253 & 6,483 \\
    Go & 317,832 & 14,242 & 14,291 \\
    \midrule
    Total &	902,690 & 47,811 & 45,229 \\
    \bottomrule
    \end{tabular}%
  \label{tab:dataset}%
  \end{table}

\begin{table}[ht]
  \centering
  \caption{Dataset Statistics}
    \begin{tabular}{lccc}
    \toprule
    \multirow{2}{*}{Language} & \multicolumn{3}{c}{Samples per partition} \\
\cmidrule{2-4}          & Train & Valid & Test \\
    \midrule
    Python & 69,527 & 11,112 & 10,565 \\
    Ruby & 11,524 & 1,075 & 1,048 \\
    JavaScript & 26,626 & 4,353 & 3,504 \\
    Go & 36,948 & 3,434 & 4,563 \\
    \midrule
    Total & 144,175 & 19,974 & 19,680 \\
    \bottomrule
    \end{tabular}%
  \label{tab:dataset_subset}%
\end{table}%

\subsection{Preprocessing}
\label{preprocessing}
We parse code snippets using Tree-Sitter\footnote{https://github.com/tree-sitter/}, an open-source parser that can parse multiple programming languages into AST. Besides, we follow the token splitting rule in~\citep{alon2019code2seq,zugner2021language,peng2021integrating} that splits each code token into sub-tokens regarding to the code name convention. For instance, the code token \texttt{sendDirectOperateCommandSet} is split into \texttt{[send, direct, operate, command, set]}. For the vocabulary of sub-tokens, we limit it with the least occurrence number of 100 in the training set, and we also restrict the max length of the token sequence to 512 after removing all punctuation. For code summarization task, we follow~\citet{zugner2021language,peng2021integrating} and remove all the anonymous functions in the javaScript dataset which can not be used for this task. For the code completion task, we randomly select a subset from the whole CSN dataset with 144,175, 19,974 and 19,680 code snippets for training, validation and testing respectively. Please see the next section
for details of the dataset used in the code completion task. For the paths, we set the max length to 32, and we make padding for the path shorter than max length while sampling nodes with equal intervals to maintain max length following~\citep{peng2021integrating}.

\subsection{Summary of the CSN Subset in Code Completion}
\label{summary_of_the_subset}
For the code completion task, we randomly select a subset from the whole CSN dataset with 144,175, 19,974 and 19,680 code snippets for training, validation and testing respectively. We show the summary of it in Table~\ref{tab:dataset_subset}.

\subsection{Baselines}
\label{baselines}
In our experiments, we set the following methods as baselines:
\begin{itemize}
    \item \textbf{Transformer:} Transformer~\citep{vaswani2017attention} is a language model based on multi-head attention, which can only model contextual information. Transformer is the basic backbone of GREAT, CodeTransformer and TPTrans.
    \item \textbf{code2seq:} Code2seq~\citep{alon2019code2seq} is an LSTM-based method that utilizes the pairwise path information in AST to model code snippets.
    \item \textbf{GREAT:} GREAT~\citep{Hellendoorn2020Global} is a Transformer-based approach that utilizes manually designed edges, such as dataflow, `computed from', `next lexical use' edges.
    \item \textbf{CodeTransformer:} CodeTransformer~\citep{zugner2021language} is a Transformer-based model that combines multiple pairwise distance relation between nodes in AST to integrate the structure information into the context sequence of code
    \item \textbf{TPTrans:} TPTrans~\citep{peng2021integrating} is a recent state-of-the-art approach for code representation learning based on Transformer and tree paths in AST, which integrate the encoding of tree paths into self-attention module by replacing the relative and absolute position embedding.
\end{itemize}

\begin{table}[htbp]
  \centering
  \caption{Number of Parameters of TPTrans and \name}
  \resizebox{0.47\textwidth}{!}{
    \begin{tabular}{lcc}
    \toprule
    \multirow{2}{*}{Model} & \multicolumn{2}{c}{Task} \\
    \cmidrule{2-3}
    & Code Summarization & Code Completion \\
    \midrule
    TPTrans & 
    113.47M & 
    101.74M \\
    
    \name-$\alpha$ & 
    178.48M 
    &
    109.87M 
    \\
    
    \name-$\beta$ & 
    118.65M 
    &
    102.73M 
    \\
    
    \name-$\gamma$ & 
    181.56M 
    &
    110.59M 
    \\
    \bottomrule
    \end{tabular}%
    }
  \label{tab:num_parameters}%
\end{table}%

\begin{table}[htbp]
  \centering
  \caption{Average One Epoch's Training Time Cost of TPTrans and \name}
  \resizebox{0.47\textwidth}{!}{
    \begin{tabular}{lcc}
    \toprule
    \multirow{2}{*}{Model} & \multicolumn{2}{c}{Task} \\
    \cmidrule{2-3}
    & Code Summarization & Code Completion \\
    \midrule
    TPTrans & 
    4h15min & 
    32min \\
    
    \name-$\alpha$ & 
    6h15min 
    &
    35min 
    \\
    
    \name-$\beta$ & 
    4h28min 
    &
    32min 
    \\
    
    \name-$\gamma$ & 
    6h44min 
    &
    35min 
    \\
    \bottomrule
    \end{tabular}%
    }
  \label{tab:training_cost}%
\end{table}%

\begin{table*}[ht]
  \centering
  \caption{Experimental results of code summarization task w/o pointer network. The bold part denotes the best results, the underlined part denotes the best results of baselines.}
    \resizebox{\textwidth}{!}{
    \begin{tabular}{lccc|ccc|ccc|ccc}
    \toprule
    \multirow{2}{*}{Model} & \multicolumn{3}{c|}{Python} & \multicolumn{3}{c|}{Ruby} & \multicolumn{3}{c|}{JavaScript} & \multicolumn{3}{c}{Go} \\
    \cmidrule{2-13}
    & Prec. & Rec.  & F1    & Prec. & Rec.  & F1    & Prec. & Rec.  & F1    & Prec. & Rec.  & F1 \\
    \midrule
    CodeTransformer (Multi.) & \underline{38.91} & 33.12 & 35.78 & 34.52 & 27.31 & 30.50 & \underline{37.21} & 29.75 & \underline{33.07} & 56.07 & 50.76 & 53.28 \\
    TPTrans (Multilingual) & 38.78 & \underline{34.72} & \underline{36.64} & \underline{38.05} & \underline{32.35} & \underline{34.97} & 36.35  & \underline{30.06} & 32.90 & \underline{56.49} & \underline{51.99} & \underline{54.15} \\
   
    \midrule
    \name-$\alpha$ & \textbf{39.26}  & 36.57 & \textbf{37.87} & \textbf{39.22} & 34.55 & 36.74 & 37.29 & 32.50 & 34.73 & \textbf{57.14} & 54.48 & 55.78 \\
    
    \name-$\beta$ & 38.34 & \textbf{37.32} & 37.82 & 38.87 & \textbf{36.07} & \textbf{37.42} & 37.35 & \textbf{34.06} & \textbf{35.63} & 56.56 & \textbf{55.14} & \textbf{55.84} \\
    
    \name-$\gamma$ & 38.50 & 36.96 & 37.71 & 38.38 & 33.99 & 36.05 & \textbf{37.72} & 32.62 & 34.98 & 56.49 & 54.30 & 55.38 \\
    
    \bottomrule
    \end{tabular}%
    }
  \label{tab:res_summarization_no_pointer}%
\end{table*}%

\subsection{Number of Parameters and Training time Cost}
We show the number of parameters of our approaches and TPTrans in Table~\ref{tab:num_parameters}. In code summarization task, compared with TPTrans, our MetaTPTrans-$\alpha$, MetaTPTrans-$\beta$ and MetaTPTrans-$\gamma$ have 57.29\%, 4.57\% and 60.01\% more parameters respectively. For code completion task, MetaTPTrans-$\alpha$, MetaTPTrans-$\beta$ and MetaTPTrans-$\gamma$ have 7.99\%, 0.97\% and 8.70\% more parameters than TPTrans respectively. In Table~\ref{tab:training_cost}, we show the average one epoch's training time cost of TPTrans and our approaches.

\subsection{Code Summarization Results w/o Pointer Network}
In our experiments on code summarization task, we apply a pointer network~\citep{vinyals2015pointer} in the decoder following~\citep{zugner2021language,peng2021integrating}. Here, we conduct an ablation study in which we remove the pointer network and the experiment results are shown in Table~\ref{tab:res_summarization_no_pointer}. We can see that, our approaches still have better performance compared with the baselines.


\end{document}